\documentclass[12pt,english]{revtex4}
\usepackage[T1]{fontenc}
\usepackage[latin1]{inputenc}
\usepackage{fancyhdr}
\usepackage{array}
\usepackage{float}
\usepackage{amsmath}
\usepackage{amssymb}

\makeatletter

\usepackage{graphicx}
\usepackage{babel}
\makeatother
\begin{document}

\title{
%Classical reactive scattering in a quantum spirit:\\A new treatment of rotational state distributions
%Classical reactive scattering in the quantum regime
Classical reactive scattering in a quantum spirit: Improving the shape of rotational state distributions 
in the quantum regime}

\author{\textit{L. Bonnet{\footnote{Corresponding author. Email: claude-laurent.bonnet@u-bordeaux.fr}}, P. Larr\'egaray, 
Ph. Halvick and J.-C. Rayez}}

\address{Institut des Sciences Mol\'eculaires, Universit\'e Bordeaux 1,
351 Cours de la Libration, 33405 Talence Cedex, France}

\begin{abstract}
For triatomic chemical reactions under single-collision conditions, we propose a new quasi-classical trajectory
(QCT) approach to rotational-state distributions of particular interest in the quantum regime where
only a few rotational states are available to the products. Our method is directly inspired 
from the amendments to be introduced in classical phase space theory (PST) in order to make it in 
exact agreement with quantum PST. The method is applied to the D$^+$ + H$_2$ and H$^+$ + D$_2$ reactions and 
the population of the rotational ground state is found to be in much closer agreement with 
the exact quantum one than the same population obtained by means of standard QCT calculations.
The impact on the whole distribution is all the stronger as the number of available states is small.
Last but not least, the shape of the distribution appears to be controlled to a large extent by three factors, 
respectively called \emph{parity, edge} 
and \emph{rotational shift} factors.
 
%Ceci a un impact sur la distribution globale d'autant plus 
%important que le nombre d'état accessibles est faible.

\end{abstract}

\maketitle

\section{Introduction}

The quasiclassical trajectory (QCT) method is widely used to study the dynamics and the kinetics of chemical reactions, 
in both the gas and condensed phases \cite{Levine,Porter,Sewell,Bon1}. Nevertheless, it may have strong intrinsic limitations, 
as regularly revealed by the comparison between its predictions and those of exact quantum mechanical (EQM) methods
\cite{Launay,Dario2,Hua0}, or high resolution experiments \cite{Yang2,Costes}. 

In this work, we focus on one of these shortcomings, namely the propensity of the standard QCT method to underestimate 
the populations of the less excited rotational states for the benefit of the most excited ones as compared 
to EQM results.
This is clearly illustrated in Fig.~\ref{fig:plot1} in the case of the D$^+$ + H$_2$
and H$^+$ + D$_2$ ion-molecule reactions (with the initial diatom in its rovibrational ground state), 
studied at collision energies sufficiently small for the product 
diatom HD to be in the vibrational ground state only. These distributions have been obtained by 
normalizing to unity the state-resolved integral cross sections (ICS) (see Eq.~\eqref{1a}) given in
references \cite{Aoiz1}  and \cite{Aoiz2} (see Fig. 6 in ref.~\cite{Aoiz1} (upper panels)
and Table II in ref.~\cite{Aoiz2}). The same conclusion could be deduced from Fig. 6 in ref.~\cite{Hua} (upper panel; 
compare the GB-QCT and QM curves) or Figs. 17 and 20 in ref.~\cite{Cza} (the comparison is between QCT and experimental 
results there). This defect is problematic in the quantum regime where only a few rotational states 
(say, up to five or six) are available to the products, as is typically the case for endoergic reactions (like H$^+$ + D$_2$) 
at collision energies slightly above threshold. 

Our goal in this paper is to find the cause and remedy of the previous limitation.
This is a key issue regarding reactions for which EQM calculations are difficult to perform. In addition, 
the QCT method has a strong interpretative power. Therefore, improving its accuracy is also important when EQM calculations 
are feasible.

Since most benchmark quantum scattering 
calculations deal with triatomic reactions, we concentrate on these processes in the following. 

The paper is organized as follows. In section II, we analyse the problem within the statistical framework of 
phase space theory (PST) \cite{PST1,Miller1,PST2,Mano,PST3,Tomas,Bon2,Bon3}. We then 
deduce the amendments to be introduced in classical PST (CPST) in order to make its predictions in 
exact agreement with those of quantum PST (QPST). These modifications are incorporated into the QCT method in section III.
In particular, parity conservation is taken into account, for the first time in the QCT approach. 
The predictions of the resulting method
are compared in section IV with EQM results and standard QCT ones in the case of the D$^+$ + H$_2$
and H$^+$ + D$_2$ reactions.
%, processes much studied in recent years. 
Section V concludes.

\section{Classical vs Quantum PST}

We consider the reaction A + BC$(n_1,j_1)$ $\longrightarrow$ AB$(n'_2,j'_2)$ + C. $n_1(n'_2)$ and $j_1 (j'_2)$ are, respectively,
the vibrational and rotational quantum numbers of BC(AB). For simplicity's sake, channel AC + B is supposed to be closed, and
both BC and AB are treated as rigid-rotor harmonic oscillators (RRHO). 
The state-resolved ICS for the previous process reads \cite{Miller1,Mano,Child}
\\
\begin{equation}
\sigma_{n'_2j'_2n_1j_1} = \frac{\pi}{k_c^2(2j_1+1)}\sum_{Jl'_2l_1}\;(2J+1)P^J_{n'_2j'_2l'_2n_1j_1l_1}.
\label{1}
\end{equation}
\\
$k_c$ is the linear momentum related to the reagent collision energy $E_c$ by $k_c = (2\mu E_c)^{1/2}/\hbar$, 
where $\mu$ is the reduced mass of A with respect to BC. $J$ is the total angular momentum quantum number and $l_1$ and $l'_2$ 
are the reagent and product orbital angular momentum quantum numbers, respectively. $P^J_{n'_2j'_2l'_2n_1j_1l_1}$ is the 
probability to start from the reagents with $(n_1,j_1,l_1)$ at $E_c$ and $J$, and reach the products with $(n'_2,j'_2,l'_2)$.

The population $P_{n'_2j'_2n_1j_1}$ of product state $(n'_2,j'_2)$ is given in terms of $\sigma_{n'_2j'_2n_1j_1}$ by
\\
\begin{equation}
P_{n'_2j'_2n_1j_1}=\frac{\sigma_{n'_2j'_2n_1j_1}}{\sum_{n'_2j'_2}\sigma_{n'_2j'_2n_1j_1}},
\label{1a}
\end{equation}
\\
where the sum in the denominator runs over all energetically accessible product states. For simplicity's sake, 
$P_{n'_2j'_2n_1j_1}$ will simply be denoted $P_{j'_2}$ in the following, as the three remaining quantum numbers will be kept at 
0 in the calculations. 

In the present case, the reaction is supposed to involve a deep well along the reaction path where the system is trapped 
enough time for complete intramolecular 
redistribution of the available energy. In other words, the reaction proceeds through an intermediate complex 
``loosing the memory'' of its initial 
conditions. Moreover, the dynamics in the entrance and exit-channels are assumed 
to be governed by isotropic long-range forces of the dispersion type. 
In such conditions, all the final states consistent
with the conservation of total energy, total angular momentum, and parity, are equally likely. 
This is the basic assumption of PST. 

Far in the reagent channel, the potential energy is given by
\\
\begin{equation}
V = \frac{1}{2} m \omega^2 (r-r_e)^2 - C_6/R^6.
\label{2}
\end{equation}
\\
$m$ is the reduced mass of BC, $\omega$ is $2\pi$ times its vibrational frequency, $r$ is the BC bond length, $r_e$ is 
its equilibrium value, $R$ is the distance between A and the center-of-mass of BC and $C_6$
is the reagent dipole-dipole dispersion coefficient which is assumed here to be the dominant one.

The internal energy of BC$(n_1,j_1)$ is given by
\\
\begin{equation}
E_{n_1j_1} = \hbar \omega \left(n_1+\frac{1}{2}\right) + \frac{\hbar^2 j_1(j_1+1)}{2mr_e^2}. 
\label{3}
\end{equation}
\\
The total energy with respect to the bottom of the reagent channel is given by
\\
\begin{equation}
E = E_c+E_{n_1j_1}
\label{4}
\end{equation}
\\ 
while its analogue in the products reads
\\
\begin{equation}
E' = E + Q
\label{5}
\end{equation} 
where $Q$ is the exoergicity.

The potential energy far in the product channel and the internal energy $E'_{n'_2j'_2}$ of AB$(n'_2,j'_2)$ are given by
the same expressions as Eqs.~\eqref{2} and~\eqref{3}, with primed coordinates 
and parameters, and $(n_1,j_1)$ replaced by $(n'_2,j'_2)$. 

$\mu'$, the reduced mass of C with respect to AB, will appear later in the developments.

\subsection{Quantum PST}
\label{II.A}

Within the previous democratic assumption of equiprobable states, $P^J_{n'_2j'_2l'_2n_1j_1l_1}$ is given by
\\
\begin{equation}
P^J_{n'_2j'_2l'_2n_1j_1l_1}=p^J_{n_1j_1l_1}\rho^J_{n'_2j'_2l'_2},
\label{4a}
\end{equation}
\\
where $p^J_{n_1j_1l_1}$ is the 
probability that the system is captured in the well when coming from the reagents with $(n_1,j_1,l_1)$ 
at $E_c$ and $J$, and $\rho^J_{n'_2j'_2l'_2}$ is the probability that it reaches the products with $(n'_2,j'_2,l'_2)$, 
subject to 
conservation of total energy, $J$ and parity. 

The capture probability $p^J_{n_1j_1l_1}$ is equal to 1 if \\\\
(i) the triangular inequality 
\begin{equation}
|j_1-l_1| \le J \le j_1+l_1
\label{6}
\end{equation}
is satisfied and \\\\
(ii) the centrifugal barrier height of the effective radial potential 
\\
\begin{equation}
V_{eff}(R) = \frac{\hbar^2 l_1(l_1+1)}{2\mu R^2} - C_6/R^6
\label{7}
\end{equation}
\\
is lower than the collision energy $E_c$. Otherwise, $p^J_{n_1j_1l_1}$ is 0. This boolean choice is meaningful
as tunneling through the centrifugal barrier is in practice negligible \cite{PST3}.

It may be shown that condition (ii) is equivalent to
\\
\begin{equation}
l_1 \le \frac{-1+\left(1+4l_M(E_c)^2/\hbar^2\right)^{1/2}}{2}
\label{12}
\end{equation}
with
\begin{equation}
l_M(E_c) = (3\mu)^{1/2}(2C_6)^{1/6}E_c^{1/3}
\label{13}
\end{equation}
\\
\cite{PST1,PST2,PST3}. $l_M(E_c)$ is in fact the maximum value of the classical orbital angular momentum consistent 
with capture at $E_c$. Note that for not too small values of 
$l_M(E_c)$, an excellent approximation of Eq.~\eqref{12} is
\\
\begin{equation}
l_1 \le \frac{l_M(E_c)}{\hbar}-\frac{1}{2}.
\label{12b}
\end{equation}
\\
From the microreversibility principle, products can only be reached with ($n'_2,j'_2,l'_2$) (at $E'$ and $J$) 
after exiting the well if the product capture probability $p^J_{n'_2j'_2l'_2}$ is equal to 1.
The necessary conditions are the same as previously, but with product notations:
\\
\begin{equation}
|j'_2-l'_2| \le J \le j'_2+l'_2
\label{6a}
\end{equation}
and
\begin{equation}
l'_2 \le \frac{-1+\left(1+4l'_M\left(E'-E'_{n'_2j'_2}\right)^2/\hbar^2\right)^{1/2}}{2}
\label{12a}
\end{equation}
with
\begin{equation}
l'_M\left(E'-E'_{n'_2j'_2}\right) = (3\mu')^{1/2}\left(2C'_6\right)^{1/6}\left(E'-E'_{n'_2j'_2}\right)^{1/3}.
\label{13a}
\end{equation}
\\
Here again, a very satisfying approximation of Eq.~\eqref{12a} is 
\\
\begin{equation}
l'_2 \le \frac{l'_M\left(E'-E'_{n'_2j'_2}\right)}{\hbar}-\frac{1}{2}.
\label{12c}
\end{equation}
\\
However, these conditions alone do not warrant that state ($n'_2,j'_2,l'_2$) is 
available when coming from the reagents with ($n_1,j_1,l_1$), $E_c$ and $J$.
An additional condition is that parity conservation is satisfied \cite{Bon3}, namely,
\\
\begin{equation}
(-1)^{(j_1+l_1)} = (-1)^{(j'_2+l'_2)}.
\label{8}
\end{equation}
Defining the \emph{parity factor} as 
\begin{equation}
\pi_{j_1l_1j'_2l'_2} = \frac{\left|(-1)^{(j_1+l_1)}+(-1)^{(j'_2+l'_2)}\right|}{2},
\label{9a}
\end{equation}
\\
parity is conserved if $\pi_{j_1l_1j'_2l'_2}=1$, non conserved if $\pi_{j_1l_1j'_2l'_2}=0$.
The boolean number
\\
\begin{equation}
\chi^J_{n'_2j'_2l'_2} = p^J_{n'_2j'_2l'_2} \pi_{j_1l_1j'_2l'_2}
\label{9}
\end{equation}
\\ 
then represents the actual contribution of $(n'_2,j'_2,l'_2)$ to the whole set of available states.
The probability to reach $(n'_2,j'_2,l'_2)$ from the well is then given by
\\
\begin{equation}
\rho^J_{n'_2j'_2l'_2}=\frac{\chi^J_{n'_2j'_2l'_2}}{\sum_{n''j''l''}\chi^J_{n''j''l''}},
\label{10}
\end{equation}
\\
where the sum in the denominator runs over all energetically accessible reactant and product states. 

To set these ideas on a simple example, let us consider, for a hypothetical process, the case $n_1=j_1=0$, $J=3$, 
implying $l_1=3$ (see Eq.~\eqref{6}), and assume that  
$p^J_{n_1j_1l_1}=p^3_{003} = 1$. For a given value of $n'_2$, the geometrical implications of the previously 
introduced constraints are depicted in Fig.~\ref{fig:plot2}. The brown ``elliptic'' curve 
is the upper bound of the partly visible grey area defined by energetic constraint~\eqref{12a}. 
The rest of this area is hidden by the yellow area, defined by both the previous constraint and 
triangular inequality~\eqref{6a}.
This area is thus bounded by the brown curve and the three blue straight lines
$J=j'_2+l'_2$, $J=j'_2-l'_2$ and $J=l'_2-j'_2$.
The allowed states $(j'_2,l'_2)$ complying with the two previous constraints,
and also parity constraint~\eqref{8}, are represented by green circles. 
Here, parity conservation forces the green states to satisfy $(-1)^{(j'_2+l'_2)}=-1$.
Prohibited states complying with the energetic constraint, triangular inequality,
but violating parity conservation, are represented by red circles. For the green states, $\chi^J_{n'_2j'_2l'_2} = 1$
(both $p^J_{n'_2j'_2l'_2}$ and $\pi_{j_1l_1j'_2l'_2}$ are equal to 1) while for the red ones, 
$\chi^J_{n'_2j'_2l'_2} = 0$ ($p^J_{n'_2j'_2l'_2}=1$ but $\pi_{j_1l_1j'_2l'_2}=0$).
The checkerboard pattern formed by the green and red circles is the direct consequence of parity conservation.

\subsection{Classical PST}
\label{II.B}

What we call the CPST estimation of $\sigma_{n'_2j'_2n_1j_1}$ is its prediction from Eq.~\eqref{1} with
$P^J_{n'_2j'_2l'_2n_1j_1l_1}$ calculated classical mechanically. 
$P^J_{n'_2j'_2l'_2n_1j_1l_1}$ is then approximated by
\\
\begin{equation}
P^J_{n'_2j'_2l'_2n_1j_1l_1}=c^J_{n_1j_1l_1}\eta^J_{n'_2j'_2l'_2},
\label{14a}
\end{equation}
\\
where $c^J_{n_1j_1l_1}$ is the 
classical probability that the system is captured in the well when coming from the reagents 
with $(n_1,j_1,l_1)$ at $E_c$ and $J$, and $\eta^J_{n'_2j'_2l'_2}$ is the probability that a trajectory
emmerging from the well contributes to the products in state $(n'_2,j'_2,l'_2)$, at $E'$ and 
$J$. 

$c^J_{n_1j_1l_1}$ is equal to 1 if triangular inequality~\eqref{6} is satisfied and if 
\\
\begin{equation}
l_1 \le l_M(E_c)/\hbar
\label{14b}
\end{equation}
\\
(see previous section). Since the difference between $l_M(E_c)/\hbar$ and the right-hand-side (RHS) of Eq.~\eqref{12} 
(see also Eq.~\eqref{12b}) is at most equal to $\sim$ 0.5,
$c^J_{n_1j_1l_1}$ and $p^J_{n_1j_1l_1}$ are generally equal.

The process being statistical, $\eta^J_{n'_2j'_2l'_2}$ is given by
\\
\begin{equation}
\eta^J_{n'_2j'_2l'_2} = \frac{F'(n'_2j'_2l'_2|E'J)}{F(EJ)+F'(E'J)}
\label{14}
\end{equation}
\\
where $F'(n'_2j'_2l'_2|E'J)$ is the flux of trajectories exiting the well towards the products 
with $E'$ and $J$ and contributing to state
$(n'_2,j'_2,l'_2)$, $F(EJ)$ is the total flux exiting the well towards the reagents with $E$ and $J$ 
and $F'(E'J)$ is the analogous flux 
towards the products. $F'(E'J)$ is the sum of $F'(n'_2j'_2l'_2|E'J)$ over $n'_2$, $j'_2$ and $l'_2$.

Convenient phase space coordinates for the mathematical formulation of the two previous
fluxes \cite{Bon2} are two radial and ten action-angle coordinates. The radial coordinates are 
the distance $R'$ between the center of mass of AB and atom C and its conjugate momentum $P'$. 
The ten action-angle coordinates are the total 
classical angular momentum $J'$, its projection $M'$ on the $z$-axis of the laboratory frame, their respective 
conjugate angles $\alpha'$ and $\beta'$, the vibrational action $n'$, the classical rotational angular momentum $j'$, 
the classical orbital angular momentum $l'$ and their respective conjugate angles $q'$, $\alpha'_j$ and $\alpha'_l$. 
These twelve coordinates form the phase space vector $\bold{\Gamma'}$. From now on, $n'$ will be expressed in $h$ unit 
and the angular momenta in $\hbar$ unit. Analogous coordinates can be used in the reagents for the
formulation of $F(EJ)$. More details on these coordinates can be found in ref.~\cite{Bon2} and references therein.

Within the previous coordinate system, $F'(n'_2j'_2l'_2|E'J)$ is given by \cite{PST2,Bon2,Bon3}
\\
\begin{equation}
F'(n'_2j'_2l'_2|E'J)=\int d\bold{\Gamma'}\delta(R'-R'_{\infty})\frac{P'}{\mu'}\Theta(P')\delta(E'-H')\delta(J'-J)
\Delta(n'-n'_2)\Delta(j'-j'_2)\Delta(l'-l'_2).
\label{15}
\end{equation}
$\Theta(x)$ is the function of Heaviside, equal to 1 if $x \ge 0$, 0 otherwise. 
$\delta$ is the Dirac delta function. $\Delta$ is the standard bin defined by 
\\
\begin{equation}
\Delta(x)=\Theta(0.5-|x|).
\label{15a}
\end{equation}
\\
Assuming that any value of $n'$ within the range $[n'_2-0.5,n'_2+0.5]$ contributes equally likely to the 
vibrational quantum state $n'_2$, with the same type of assignment for $j'$ and $l'$, is called 
the \emph{standard binning} (SB) procedure.
$R'_{\infty}$ is an infinitely large value of $R'$. The classical Hamiltonian $H'$ in the product channel reads
\\
\begin{equation}
H' = \frac{{P'}^2}{2\mu'}+{V^C_{eff}}'(R')+E^C_{n'j'}
\label{16}
\end{equation}
with
\begin{equation}
{V^C_{eff}}'(R') = \frac{\hbar^2 l'^2}{2\mu' {R'}^2} - C'_6/{R'}^6
\label{17}
\end{equation}
and
\begin{equation}
{E'}^C_{n'j'} = \hbar \omega' \left(n'+\frac{1}{2}\right) + \frac{\hbar^2 {j'}^2}{2m'{r'_e}^2}. 
\label{18}
\end{equation}
\\
%$\mu'$, $m'$, $C'_6$, $r'_e$ and $\omega'$ are the counterpart 
%of the previously defined similar unprimed quantities. 
The superscript $C$ in ${V^C_{eff}}'(R')$ and $E^C_{n'j'}$ is to recall that these energies are classical, contrary to
the quantum mechanical energies $V_{eff}(R)$ and $E_{n_1j_1}$ (see Eqs.~\eqref{7} and~\eqref{3}). 

Following refs.~\cite{PST2,Bon2,Bon3}, we arrive after a few steps of algebra, at
\\
\begin{equation}
F'(n'_2j'_2l'_2|E'J)\propto\int dn' dj' dl'\Delta(n'-n'_2)\Delta(j'-j'_2)\Delta(l'-l'_2).
\label{19}
\end{equation}
\\
The upper limit of $l'$ is 
\begin{equation}
l'=l'_M(E'-E'^C_{n'j'}),
\label{19b}
\end{equation}
\\
and $j'$ and $l'$ satisfy the triangular inequality
$|j'-l'| \le J \le j'+l'$.

\subsection{Modifiying CPST so as to make it equivalent to QPST}
\label{II.C}

Comparing Eqs.~\eqref{4a} and~\eqref{14a}, the modification to perform in CPST in order to make it in 
exact agreement with QPST should be such that, within the modified CPST,
\\ 
\begin{equation}
c^J_{n_1j_1l_1} = p^J_{n_1j_1l_1}
\label{20a}
\end{equation}
and 
\begin{equation}
\eta^J_{n'_2j'_2l'_2} = \rho^J_{n'_2j'_2l'_2},
\label{20}
\end{equation}
or equivalently,
\begin{equation}
F'(n'_2j'_2l'_2|E'J)\propto\chi^J_{n'_2j'_2l'_2}
\label{19a}
\end{equation}
\\
(see Eqs.~\eqref{10} and~\eqref{14}). These conditions will be fulfilled if the following method
%, later shown to be straightforwardly implementable in a QCT code, 
is used:\\\\
(i) We replace the classical upper bound of $l_1$ (Eq.~\eqref{14b}) by its quantum analogue (Eq.~\eqref{12}),
thus making identity~\eqref{20a} satisfied.
\\\\
(ii) The $\Delta$ functions in Eq.~\eqref{19} are replaced by Gaussian functions of the type 
\\
\begin{equation}
G(x) = \frac{e^{-x^2/\epsilon^2}}{\pi^{1/2} \epsilon},
\label{21}
\end{equation}
\\
normalized to unity, with $\epsilon$ tending to 0$^+$. 
%enough for the Gaussians to have a full-width-at-half-maximum (FWHM) negligible as compared 
%to the unit distance between nearest Gaussian centers in the action space $(n',j',l')$. 
Therefore, these Gaussians are equivalent to Dirac delta functions. 
%This procedure is called Gaussian binning (GB) \cite{Bon1,Aoiz1,Aoiz2,Hua,Bon3,Part1,Part2,Part3,Part4,Part5}.
We may then integrate over $n'$ in Eq.~\eqref{19}, thereby getting 
\\
\begin{equation}
F'(n'_2j'_2l'_2|E'J)\propto\int dj' dl'G(j'-j'_2)G(l'-l'_2).
\label{22}
\end{equation}
\\
$n'$ being now equal to $n'_2$, the classical upper bound of $l'$ is $l'_M(E'-E'^C_{n'_2j'})$ (see Eq.~\eqref{19b}).
\\\\
(iii) We artificially impose in Eq.~\eqref{22}
\\ 
\begin{equation}
l' = \frac{-1+\left(1+4l'_M\left(E'-E'_{n'_2j'}\right)^2/\hbar^2\right)^{1/2}}{2}
\label{22a}
\end{equation}
\\
as an upper limit for $l'$. $l'_M(E'-E'_{n'_2j'})$ is given by Eq.~\eqref{13a} and 
$E'_{n'_2j'}$ by Eq.~\eqref{3} with \emph{ad-hoc} product parameters. 
This quantum boundary is compared with the classical boundary $l'_M(E'-E'^C_{n'_2j'})$ in Fig.~\ref{fig:plot2a} for
H$^+$ + D$_2$(0,0) at $E_c=$ 102 meV, leading to D$^+$ and HD in the vibrational ground state only. 
The values of the parameters necessary to obtain these curves can be found in ref.~\cite{Bon2}.
When carefully looking at Eqs.~\eqref{22a},~\eqref{13a} and~\eqref{3} (with product parameters) on the one hand, 
and Eqs.~\eqref{19b} and~\eqref{18} on the other hand, one arrives at the conclusion that one goes from the blue
to the red curve in Fig.~\ref{fig:plot2a} by replacing $j'$ by 
\\
\begin{equation}
j'_q(j')= \frac{-1+\left(1+4j'^2\right)^{1/2}}{2}
\label{22x}
\end{equation}
and $l'$ by
\begin{equation}
l'_q(l')= \frac{-1+\left(1+4l'^2\right)^{1/2}}{2},
\label{22y}
\end{equation}
\\
approximated by $j'-1/2$ and $l'-1/2$, respectively, for not too small values of $j'$ or $l'$. 
This observation will be useful later in this work.
The quantum limit is thus roughly shifted in by $\sim$ 0.5 with respect to the classical one along both 
the $j'$ and $l'$ axes. In the present case, this shift plays a major role: the green states $(j'_2,l'_2)$ 
in Fig.~\ref{fig:plot2a} are indeed
classically available, but quantally prohibited, thus implying that the rotational state population $P_3$ is 0 in QPST, but not in CPST.
\\\\
(iv) Eq.~\eqref{22} is now modified according to
\\
\begin{equation}
F'(n'_2j'_2l'_2|E'J)\propto\pi_{j_1l_1j'_2l'_2} \; e^J_{j'_2l'_2}\int dj' dl'G(j'-j'_2)G(l'-l'_2)\kappa_(j'_2,l'_2).
\label{23}
\end{equation}
\\
$\kappa_(j'_2,l'_2)$ is equal to 1 if $(j'_2,l'_2)$ lies below the red curve in Fig.~\ref{fig:plot2a}, 0 if it lies 
between the red and blue curves. Hence, $\kappa_(j'_2,l'_2)$ makes 0 the contribution of the state consistent 
with the classical capture but not the quantum one. As the basic reason responsible for the differences between the 
classical and quantum capture limits is that the rotational energy associated with a given angular 
momentum is slightly larger in classical than in quantum mechanics, we shall call $\kappa_(j'_2,l'_2)$ the 
\emph{rotational shift factor}. Since the Gaussians are supposed to be infinitely narrow, one may rewrite Eq.~\eqref{23} as
\\
\begin{equation}
F'(n'_2j'_2l'_2|E'J)\propto\pi_{j_1l_1j'_2l'_2} \; e^J_{j'_2l'_2}\int dj' dl'G(j'-j'_2)G(l'-l'_2)\kappa_(j',l'),
\label{23a}
\end{equation}
\\
where $\kappa_(j',l')$ is as previously defined, but for real values of $j'$ and $l'$. 
This expression will appear to be useful in QCT calculations. 

As previously discussed in section~\ref{II.A}, the parity factor $\pi_{j_1l_1j'_2l'_2}$ 
makes the flux $F'(n'_2j'_2l'_2|E'J)$ equal to 0 when parity conservation is not respected, but leaves unchanged the rest of the RHS
of Eqs.~\eqref{23} and~\eqref{23a} in the contrary case.

The remaining factor is defined by
\\
\begin{equation}
e^J_{j'_2l'_2}=2^{\left(\delta_{J,l'_2+j'_2}+\delta_{J,l'_2-j'_2}+\delta_{J,j'_2-l'_2}\right)}
\label{24}
\end{equation}
\\
where $\delta_{m,n}$ is the Kronecker symbol, equal to 1 if the two integers $m$ and $n$ are identical, 0 otherwise.
We call it the \emph{edge factor} for reasons that will appear obvious further below.
For the example of section~\ref{II.A}, 
the bidimensional Gaussians $G(j'-j'_2)G(l'-l'_2)$ centered at the available states $(j'_2,l'_2)$ 
(green circles in Fig.~\ref{fig:plot2}) are schematically 
represented in Fig.~\ref{fig:plot3}. 
A zoom of the Gaussian centered at (3,0) is also shown. Colored discs represent 
the areas where Gaussians take significant values. These areas have 
been arbitrarily increased for clarity's sake, as we have previously assumed that $\epsilon$ tends to 0$^+$, thereby 
implying that these areas tend to 0. 
For the magenta, orange and green Gaussians, $e^J_{j'_2l'_2}=$ 4, 2 and 1, respectively. Note that the substitution 
of the quantum boundaries of angular momenta to the classical values by means of the rotational shift 
factor $\kappa(j',l')$ makes the brown upper bounds in 
Figs.~\ref{fig:plot2} and~\ref{fig:plot3} rigorously identical. Otherwise, the upper bound in 
Fig.~\ref{fig:plot3} would be roughly shifted out by one half (see Fig.~\ref{fig:plot2a}).

%Note that the brown upper bound in Fig.~\ref{fig:plot2} is slightly below its red analog in 
%Fig.~\ref{fig:plot3} (though this cannot be seen in the figures). 
%The former is indeed given by $l'\approx l'_M(E'-E'_{n'_2j'})-1/2$ (see Eq.~\eqref{12c}) 
%while the latter is given by $l'=l'_M(E'-E'^C_{n'_2j'})$, with $E'_{n'_2j'}$ slightly
%larger than $E'^C_{n'_2j'}$ (compare Eqs.~\eqref{3} and \eqref{18}). In principle, one may thus 
%have classically available states $(j'_2,l'_2)$ that are quantum mechanically forbiden 
%(if $l'_M(E'-E'_{n'_2j'_2})-1/2 \le l'_2 \le l'_M(E'-E'^C_{n'_2j'_2})$), though their number is expected to be
%negligible as compared to the total number of available states.
% (in both classical and quantum mechanics). 

We are now in a position to perform the integration over $j'$ and $l'$ in Eq.~\eqref{23} for the example at hand. 
For $(j'_2,l'_2)$ corresponding to the two magenta Gaussians
in Fig.~\ref{fig:plot3}, only one fourth of the Gaussians 
lie within the area imposed by the triangular inequality. This is clearly seen in the zoom of the Gaussian centered at (3,0). 
Hence, it is 
clear that their integration over $j'$ and $l'$ leads to one fourth. But for these Gaussians, $e^J_{j'_2l'_2}=4$. 
Consequently, $F'(n'_2j'_2l'_2|E'J)\propto1$. For the four orange Gaussians, 
half of the Gaussians lie within the previous area, but $e^J_{j'_2l'_2}=2$. The two remaining
green Gaussians lie entirely within the area imposed by the triangular inequality, and $e^J_{j'_2l'_2}=1$. 
As a consequence, the edge factor makes $F'(n'_2j'_2l'_2|E'J)\propto1$ for the eight Gaussians represented in Fig.~\ref{fig:plot2}, i.e., 
for the green states in Fig.~\ref{fig:plot2}.  
Moreover, $F'(n'_2j'_2l'_2|E'J)$ is 0 for the red states in Fig.~\ref{fig:plot2}, due to the $\pi_{j_1l_1j'_2l'_2}$ 
factor in Eq.~\eqref{23}.
As a consequence, $F'(n'_2j'_2l'_2|E'J)$, such as given by Eq.~\eqref{23}, is proportional to $\chi^J_{n'_2j'_2l'_2}$
and Eq.~\eqref{20} is then satisfied (see also Eqs.~\eqref{10} and~\eqref{14}). 

We now know that quantizing the vibrational, rotational and orbital motions by means of infinitely narrow Gaussians, 
and including the parity, edge and rotational shift factors in CPST makes it in exact agreement with QPST \cite{Note1}. 

In order to illustrate this finding, the predictions of $P_{j'_2}$ obtained by means of QPST and the modified CPST are
represented in Fig.~\ref{fig:plot4} for D$^+$ + H$_2$(0,0) at $E_c = 100$ meV (red columns and blue diamonds, respectively). 
$\epsilon$ was kept at 0.06, a value for the which the full-width-at-half-maximum (FWHM) of the Gaussians is equal to 
10$\%$. When used in QCT calculations, this procedure is commonly termed \emph{Gaussian binning} (GB)
\cite{Bon1,Aoiz1,Aoiz2,Hua,Cza,Part1,Part2,Part3,Part4,Part5}. 
Below this value, the Monte-Carlo estimation of Eq.~\eqref{23a} starts getting harder to converge in a a few 
minutes, the usual amount of time required for QPST or CPST calculations.
%The substitution of the $\Delta$ function by a $\sim$ 10$\%$ wide Gaussian is called the Gaussian binning (GB) procedure
%\cite{Bon1,Aoiz1,Aoiz2,Hua,Bon3,Part1,Part2,Part3,Part4,Part5}.  
The values of the remaining parameters necessary to 
perform the PST calculations can be found in ref.~\cite{Bon2}.
As a matter of fact, the QPST and modified CPST predictions appear to be in excellent agreement.
Note, however, that the modified CPST populations are slightly lower than the QPST ones. This is due to 
the fact that for the set of available quantum states very close to the brown upper bound in Fig.~\ref{fig:plot2} 
or~\ref{fig:plot3}, the Gaussians partly overlap the forbiden region in the $(j', l')$ plane (unless
$\epsilon$ tends to 0$^+$). Their whole contribution is thus less than the number of previous states. But since the latter
do not represent a large part of the whole set of available states, the above mentioned underestimation has only
a minor impact on the final results.

\section{New implementation of the QCT method}

The constraints previously introduced in CPST will now be included into the QCT approach, such as implemented in
section 2 of ref.\cite{Bon1}, or ref.\cite{Bon2}. 

In this method, action-angle coordinates such as those
discussed in section~\ref{II.B} are used to generate the initial conditions corresponding to the collision energy
$E_c$ and the quantum numbers
$n_1$, $j_1$, $l_1$ and $J$ appearing in the general expression~\eqref{1} of the state-resolved ICS. 
%These coordinates have mainly been introduced in molecular scattering theory by Miller, 
%in his pioneering semiclassical 
%developments of the early seventies \cite{Miller2,Miller3}.

Like in most QCT approaches, trajectories are run 
from a large initial distance $R_i$, such that A and BC do not interact, with the radial momentum
\\
\begin{equation}
P=-\left[2\mu\left(E_c-\frac{l_1(l_1+1)}{2\mu R^2}\right)\right]^{1/2}
\label{25}
\end{equation}
\\
(for not too large values of $R_i$, the centrifugal energy may not be negligible as compared to $E_c$). 
The ten remaining action-angle coordinates to select are the total 
classical angular momentum, kept at $J$, its projection on the $z$-axis, kept at any value between $-J$ and $J$, 
their respective conjugate angles $\alpha$ and $\beta$, arbitrary, the vibrational action $n$, taken
at $n_1$, the classical rotational angular momentum $j$ and the classical orbital angular momentum $l$, respectively 
kept at $j_q(j_1)$ and $l_q(l_1)$ (see Eqs.~\eqref{22x} and~\eqref{22y}), 
and their conjugate angles $q$, $\alpha_j$ and $\alpha_l$, randomly chosen between 0 and $2\pi$. 
The last three angles at time 0 are collectively denoted $\bold{q_1}$. Note that taking $l$ at $l_q(l_1)$
is equivalent to step (i) of the method proposed in section~\ref{II.C}.

To avoid any numerical instability, trajectories are run in Cartesian coordinates. The passage from 
$R$, $P$ and the ten action-angle coordinates, to Cartesian ones, can be found in refs.~\cite{Bon2,Miller2} (the 
similar transformation for polyatomic processes is given in ref.~\cite{Bon4}). The product vibrational action, 
rotational angular momentum and orbital angular momentum are respectively denoted $n'({\bold{q_1}})$, $j'({\bold{q_1}})$ and 
$l'({\bold{q_1}})$. 

Following the developments of section 2 in ref.\cite{Bon1}, the QCT expression of $P^J_{n'_2j'_2l'_2n_1j_1l_1}$, 
including the modifications previously introduced in CPST (see steps (ii)-(iv) of the method proposed in section~\ref{II.C})
reads  
\\ 
\begin{equation}
P^J_{n'_2j'_2l'_2n_1j_1l_1} = \frac{Q^J_{n'_2j'_2l'_2n_1j_1l_1}}
{\sum_{n_2j_2l_2}Q^J_{n_2j_2l_2n_1j_1l_1}+\sum_{n'_2j'_2l'_2}Q^J_{n'_2j'_2l'_2n_1j_1l_1}}
\label{26}
\end{equation}
with
\begin{equation}
Q^J_{n'_2j'_2l'_2n_1j_1l_1} = \pi_{j_1l_1j'_2l'_2} e^J_{j'_2l'_2}
 \int_{D_R} \; d{\bold{q_1}} \; G_{n'}\;G_{j'}\;G_{l'}\;\kappa_{j'l'}, 
\label{27}
\end{equation}
\\
\begin{equation}
G_{n'} = G(n'({\bold{q_1}})-n'_2),
\label{28}
\end{equation}
\\
\begin{equation}
G_{j'} = G(j'({\bold{q_1}})-j'_2)
\label{29}
\end{equation}
\\
\begin{equation}
G_{l'} = G(l'({\bold{q_1}})-l'_2)
\label{30}
\end{equation}
and
\begin{equation}
\kappa_{j'l'} = \kappa(j'({\bold{q_1}}),l'({\bold{q_1}})).
\label{30a}
\end{equation}
\\
The practical calculation of the rotational shift factor $\kappa_{j'l'}$ is discussed further below.
In Eq.~\eqref{27}, integration is made over the domain $D_R$ of initial angles leading to reactive trajectories. 
$Q^J_{n_2j_2l_2n_1j_1l_1}$, in the first sum of the denominator of Eq.~\eqref{26}, is given by a similar expression involving 
non reactive trajectories. Note that $P^J_{n'_2j'_2l'_2n_1j_1l_1}$ is unitary, i.e., its sum
over all energetically accessible reactant and product states leads to 1, as it should be. 
%We shall come back to this 
%normalization issue in the next section. 

Eq.~\eqref{27}, however, has a serious defect. The product of three narrow Gaussians in its integrand 
makes the calculation very heavy. Using, for instance, Gaussians with
FWHM of 10 $\%$, only $\sim$ 0.1 $\%$ of the trajectories do actually contribute to 
$P^J_{n'_2j'_2l'_2n_1j_1l_1}$ \cite{Part5}. Therefore, the calculations presented in the next section have been 
performed by means of the GB procedure for the vibration motion only, while the SB procedure 
has been used for the rotational and orbital angular motions. In this case, 
\\
\begin{equation}
Q^J_{n'_2j'_2l'_2n_1j_1l_1} = \pi_{j_1l_1j'_2l'_2} e^J_{j'_2l'_2}
 \int_{D_R} \; d{\bold{q_1}} \; G_{n'}\;\Delta_{j'}\;\Delta_{l'}\;\kappa_{j'l'} 
\label{31}
\end{equation}
with 
\begin{equation}
\Delta_{j'} = \Delta(j'({\bold{q_1}})-j'_2)
\label{32}
\end{equation}
and
\begin{equation}
\Delta_{l'} = \Delta(l'({\bold{q_1}})-l'_2)
\label{33}
\end{equation}
\\
(see Eq.~\eqref{15a}). This increases the number of efficient trajectories to $\sim$ 10 $\%$,
a quite acceptable value. 

The calculation of $\kappa_{j'l'}$ is as follows. We have seen in the previous section that one goes from the blue
to the red curve in Fig.~\ref{fig:plot2a} by replacing $j'$ and $l'$ by $j'_q(j')$ and $l'_q(l')$, respectively,
in the analytical expression of the blue curve. 
$\kappa_{j'l'}$ was then kept at 1 below the red curve, and 0 between the red and blue curves. Note that $n'$ was set
equal to $n'_2$ exactly. But it should be clear from Eqs.~\eqref{12a} and~\eqref{13a} that the red curve is also 
an upper bound of the area defined by the previous expressions and any value of $n'$ larger than $n'_2$ 
($E'_{n'j'}$ is indeed an increasing function of $n'$, and $l'_M(E'-E'_{n'j'})$ a decreasing one).
Calling, respectively, $\bar{l}'$ and $\bar{j}'$ the rounded values of $l'$ and $j'$, the idea is thus to find the 
maximum $j'_M(\bar{l}')$ of $j'_q(j')$ for all the values of $l'$ corresponding to the same $\bar{l}'$, 
and analogously, the maximum $l'_M(\bar{j}')$ of $l'_q(l')$ for all the values of $j'$ corresponding to the same 
$\bar{j}'$. From the previous remark on $n'$, all the trajectories leading to 
$n'\ge n'_2$ can be taken into account. $\kappa_{j'l'}$ is then kept at 1 if both $j'$ and $l'$ are lower than $j'_M(\bar{l}')$
and $l'_M(\bar{j}')$, respectively, 0 otherwise. Finally, $P_{j'_2}$ is kept at 0 for any value of $j'_2$ larger than the maximum 
of the $j'_M(\bar{l}')$'s. 

$Q^J_{n'_2j'_2l'_2n_1j_1l_1}$ was numerically estimated by randomly selecting 
$\bold{q_1}$ and summing the integrand of Eq.~\eqref{31}. The number of trajectories run for each value of $J$ 
was chosen to be proportional to $2J+1$.

\section{Application to the D$^+$ + H$_2$ and H$^+$ + D$_2$ reactions}

Batches of $\sim 10^5$ trajectories were run for (a) D$^+$ + H$_2$(0,0) at $E_c=100$ meV, (b) the same process at $E_c=190$ meV,
and (c) H$^+$ + D$_2$(0,0) at $E_c=102$ meV. They were run on the same potential energy surface (PES) as in 
refs.~\cite{Aoiz1}  and \cite{Aoiz2}, i.e., the PES of Aguado \emph{et al.} \cite{Aguado}.
Eqs.~\eqref{1},~\eqref{1a} and~\eqref{31} with $\epsilon=0.1$ lead to the blue diamonds in Fig.~\ref{fig:plot5}, 
to be compared with the red circles and the green squares in Fig.~\ref{fig:plot1}. Note that the latter have been 
found by using GB for $n'$, and SB for $j'_q(j')$ \cite{Aoiz1,Aoiz2}. As a matter of fact, the present
QCT method leads to a much better prediction of $P_0$ than the standard QCT method. The former approach systematically 
enhances $P_0$ by a factor of $\sim 2$ compared to the latter. Consequently, the remaining populations are slightly 
decreased (except $P_1$ which is almost unchanged), thereby improving to some extent the overall shape of the rotational 
state distribution, especially for H$^+$ + D$_2$(0,0) in which only three states are available. 

On the other hand, the modified (standard) QCT total ICS is found to be equal to 35.2 (28.2), 20.2 (22.8) and 15.4 (15.5) \AA$^2$ 
for reactions (a), (b) and (c), respectively, against 33.7, 27.6 and 23.6 \AA$^2$ from EQM calculations. Therefore, the modified 
QCT method only improves the ICS for reaction (a). For the remaining processes, both QCT methods tend to underestimate the ICS,
a known defect studied in refs.\cite{Bon1} and~\cite{Part3}, which precise origins remain to be clearly established.

\section{Conclusion}

In the quantum regime where only a few rotational states are available to reaction products, 
the standard quasi-classical trajectory (QCT) method often underestimates the population of the less excited rotational states
for the benefit of the most excited ones.

In the present work, we have analized the reasons for this underestimation in the light of the statistical 
phase space theory (PST) of chemical reactions. We have found three main sources of disagreement between classical PST (CPST) 
and quantum PST (QPST), and have introduced three related corrections making the modified CPST in agreement with QPST.
These corrections consist in multiplying the final phase space states by a product of three factors, respectively called 
\emph{parity, edge} and \emph{rotational shift factors}, controlling to a large extent the shape 
of rotational state distributions. 

These corrections have then been implemented in the QCT approach. For the ion-molecule reactions D$^+$ + H$_2$ and H$^+$ + D$_2$,
the modified QCT prediction of the rotational ground state population turns out to be in much better agreement with the exact quantum 
mechanical one than the standard QCT prediction. 
The impact on the whole distribution is all the stronger as the number of available states is small.

The next steps are twofold: (i) checking that 
this improvement is general in the quantum regime by studying more processes and (ii) using this new QCT method to
rationalize the shape of rotational state distributions in terms of energetic and mechanical parameters like collision energy, 
exoergicity, vibrational frequencies, atomic masses, etc... 

Last but not least, this work is one more illustration of the fact that the QCT method may be much more accurate than 
expected two decades ago, provided that some quantum constraints are added to it in the light of the quantum and 
semiclassical formalisms. This is good news for polyatomic reactions which can rather easily be studied by the 
QCT method \cite{Cza,Joa}, while hardly by EQM approaches.

\newpage

\section*{Figures captions}

\begin{figure}[H]
\begin{center}\includegraphics[%
  clip,
  angle=0,
  origin=c]{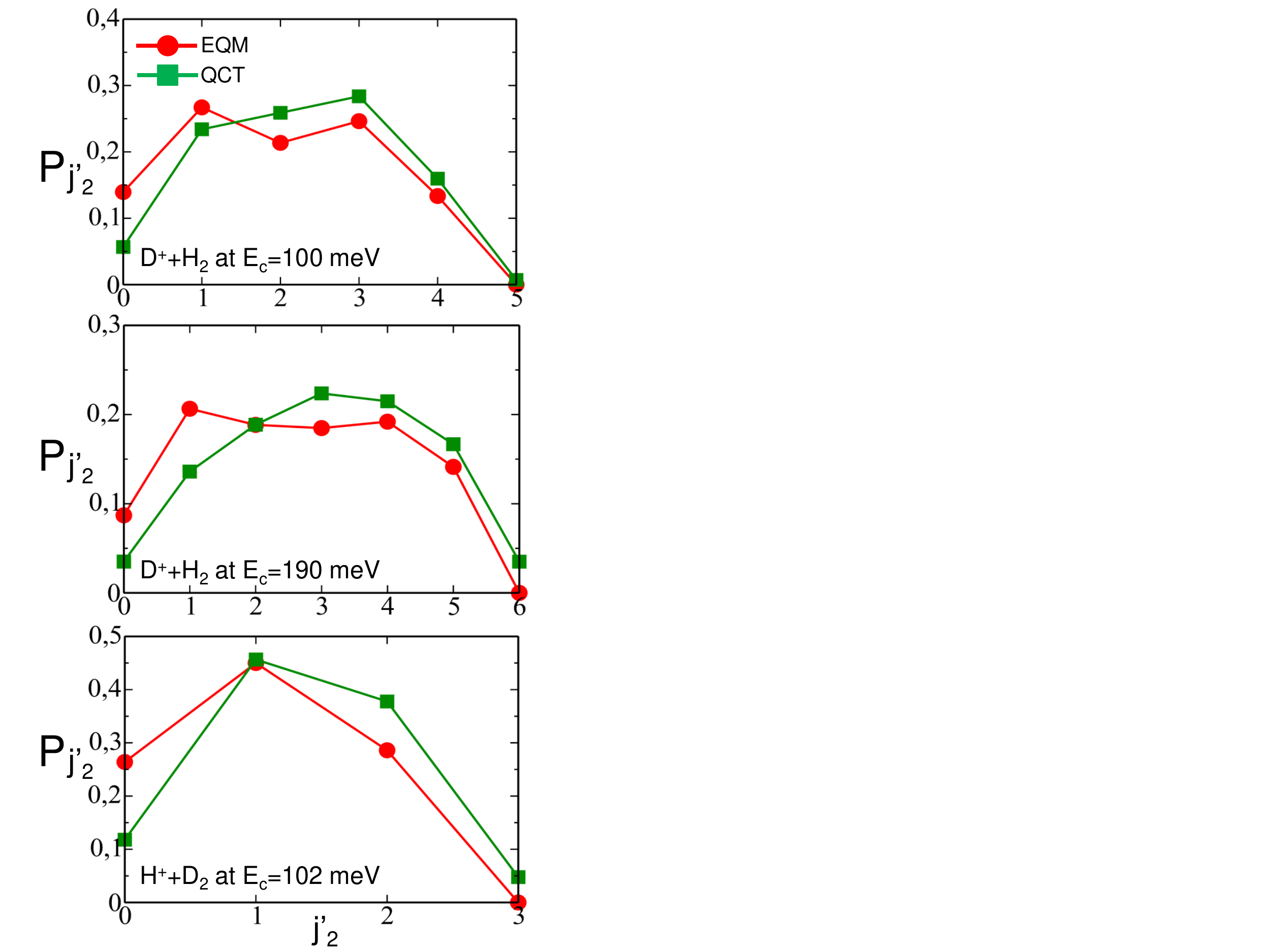}\end{center}

\caption{\label{fig:plot1}}
\end{figure}

\begin{figure}[H]
\begin{center}\includegraphics[%
  clip,
  angle=0,
  origin=c]{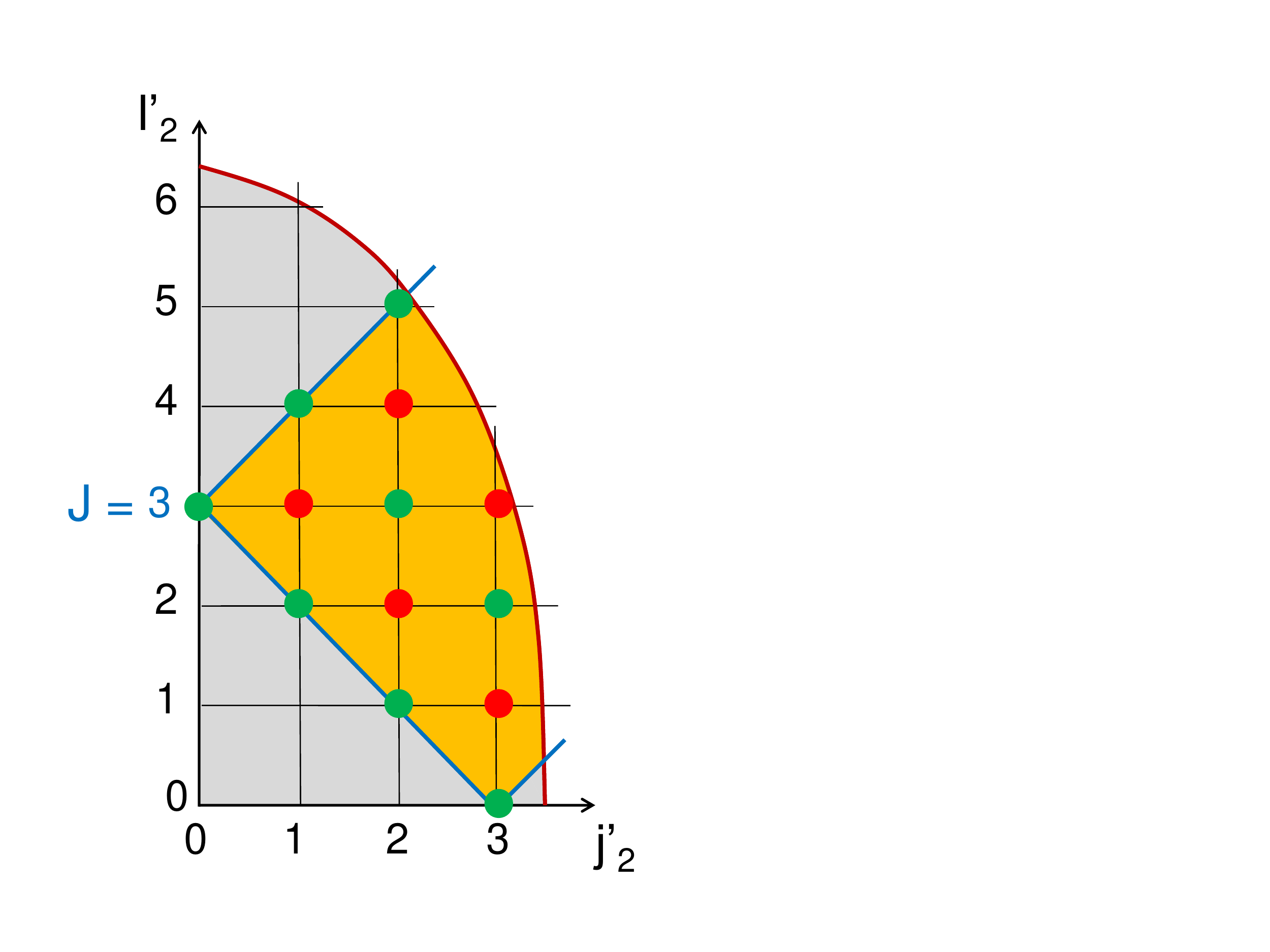}\end{center}

\caption{\label{fig:plot2}}
\end{figure}

\begin{figure}[H]
\begin{center}\includegraphics[%
  clip,
  angle=0,
  origin=c]{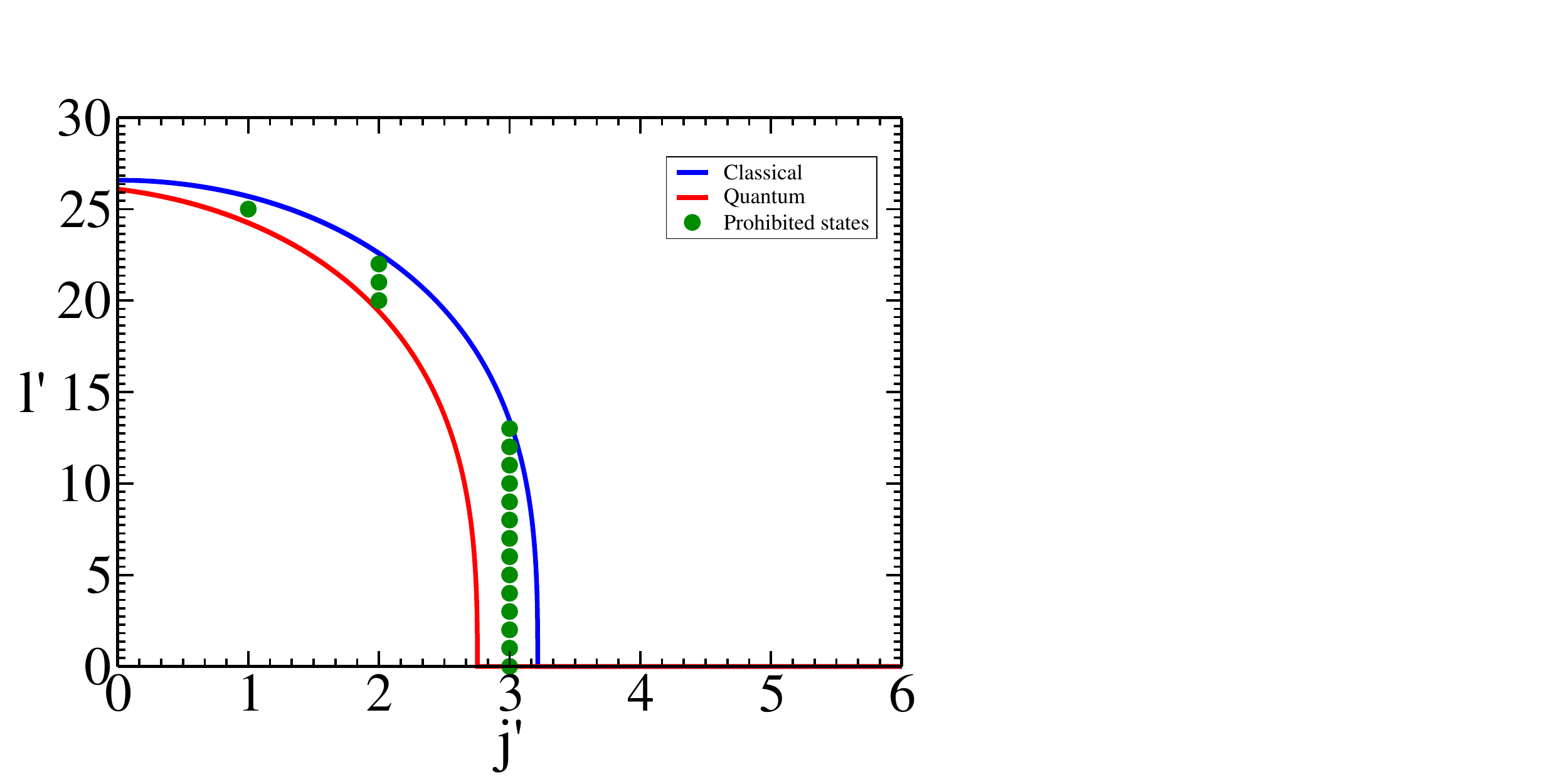}\end{center}

\caption{\label{fig:plot2a}}
\end{figure}

\begin{figure}[H]
\begin{center}\includegraphics[%
  clip,
  angle=0,
  origin=c]{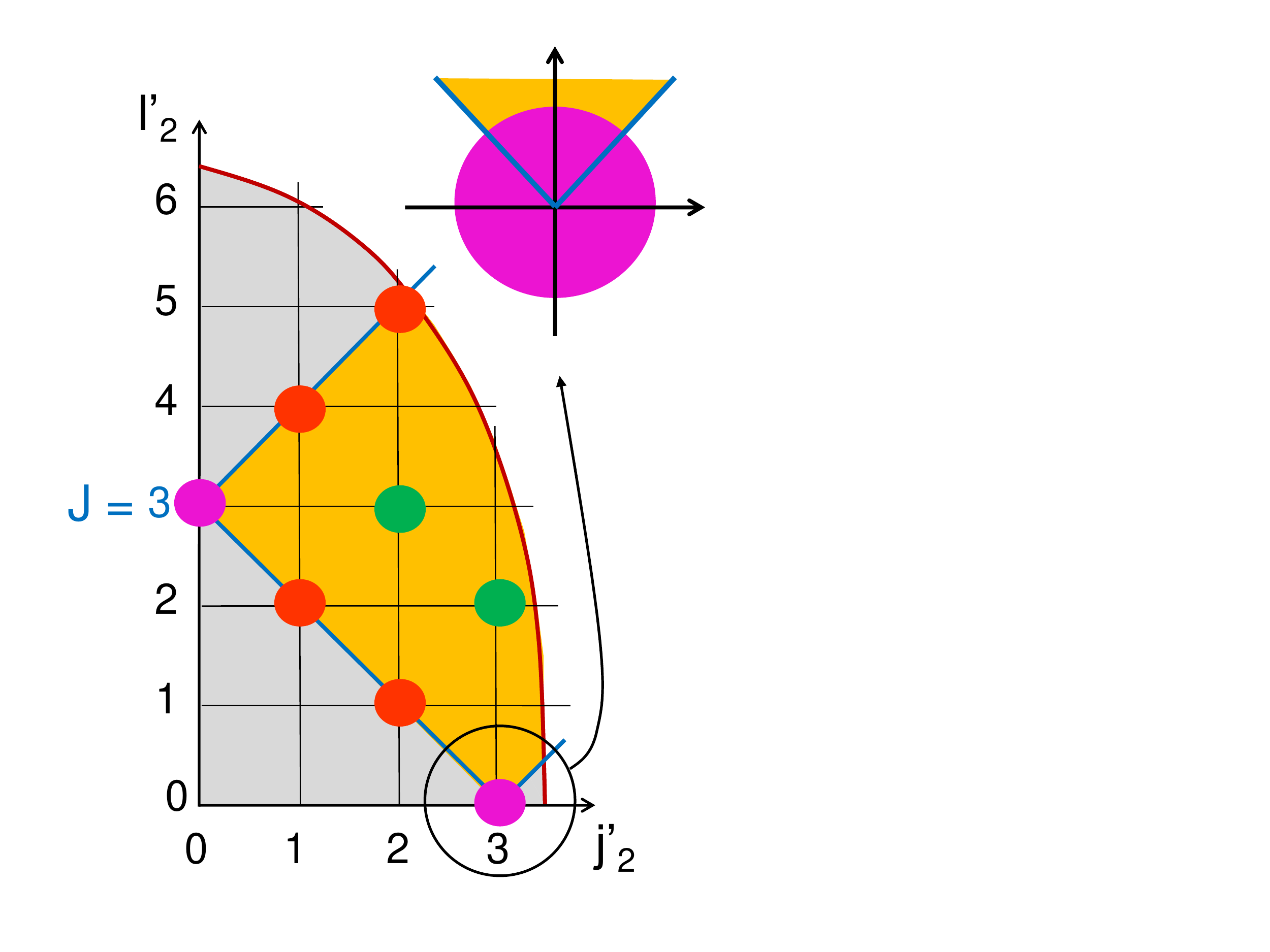}\end{center}

\caption{\label{fig:plot3}}
\end{figure}

\begin{figure}[H]
\begin{center}\includegraphics[%
  clip,
  angle=0,
  origin=c]{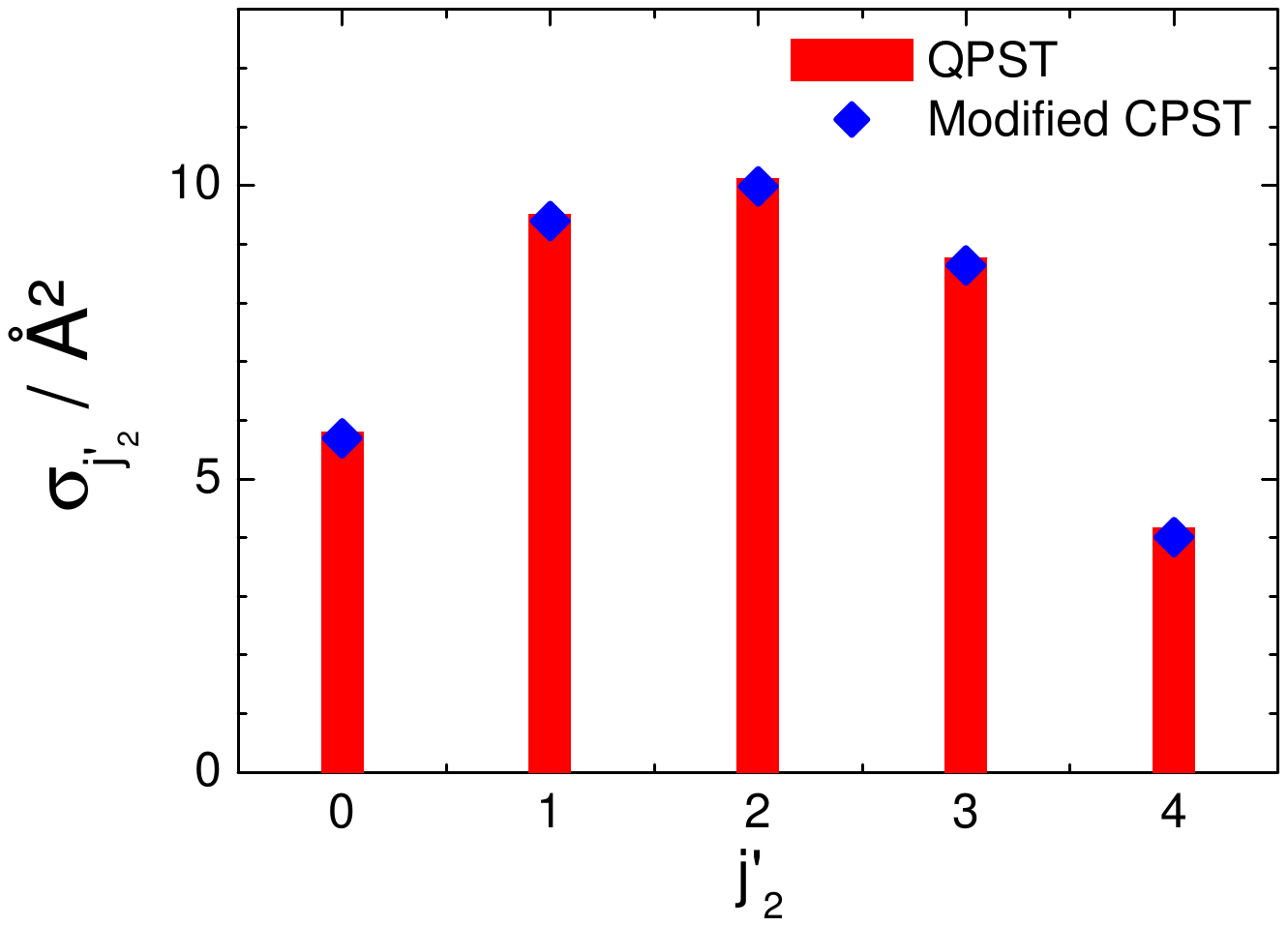}\end{center}

\caption{\label{fig:plot4}}
\end{figure}

\begin{figure}[H]
\begin{center}\includegraphics[%
  clip,
  angle=0,
  origin=c]{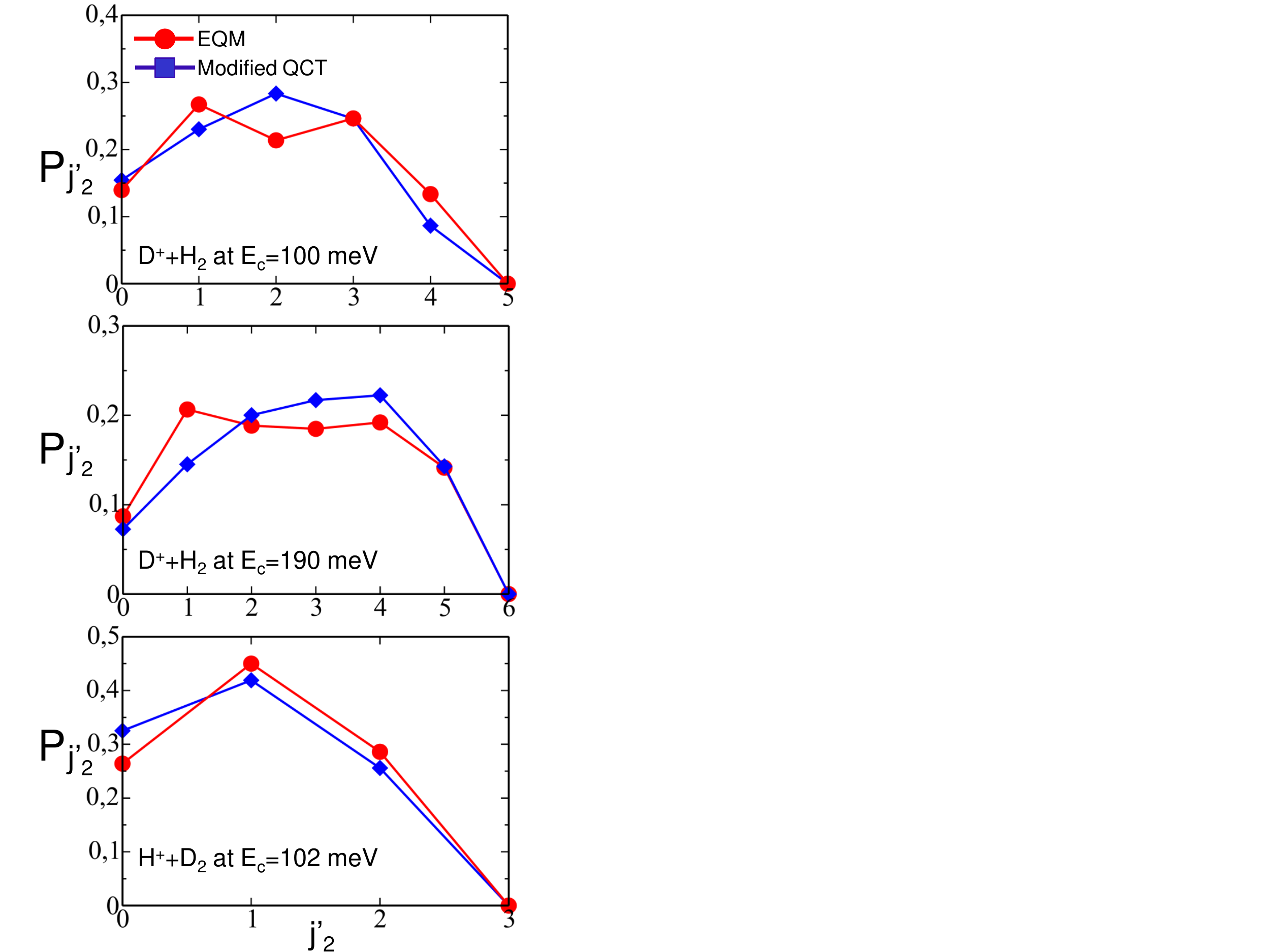}\end{center}

\caption{\label{fig:plot5}}
\end{figure}

Fig. \ref{fig:plot1}: Distribution of the rotational quantum number $j'_2$ of HD formed from the reactions D$^+$ + H$_2$
and H$^+$ + D$_2$. The initial diatom is in its rovibrational ground state. The values of the collision energy $E_c$ are 
indicated. These are sufficiently low for HD to be in its vibrational ground state only.
Red circles and green squares correspond to exact quantum mechanical (EQM) and quasiclassical trajectory (QCT) results, 
respectively. These distributions have been obtained by 
normalizing to unity the state-resolved ICSs given in references \cite{Aoiz1}  and \cite{Aoiz2}.
$P_0$, as well as $P_1$ for D$^+$ + H$_2$, appear to be underestimated by QCT compared to EQM. 
Consequently, the remaining populations tend to be overestimated by QCT. \\

Fig. \ref{fig:plot2}: Checkerboard pattern formed by allowed (green circles) and prohibited states (red circles) 
in the $(j'_2,l'_2)$ plane for a hypothetical process, a given value of $n'_2$ and $J=3$. The partly visible grey 
area represents the energetically available region while the yellow area is the part of the previous region 
allowed by triangular inequality. Hence, the represented states satisfy energy and total angular momentum
requirements. Among them, however, only the green states comply with parity conservation
(see text for more details). \\

Fig. \ref{fig:plot2a}: Quantum upper bound (red curve) compared with its classical analogue (blue curve)
in the products of the reaction H$^+$ + D$_2$(0,0) at $E_c=$ 102 meV.
The quantum limit is roughly shifted in by $\sim$ 0.5 with respect to the classical one along both 
the $j'$ and $l'$ axes. The green states are classically available, but quantally prohibited. 
$P_3$ is thus 0 in QPST, but not in CPST.  \\

Fig. \ref{fig:plot3}: For the process considered in Fig. \ref{fig:plot2}, colored discs represent the areas where, 
Gaussians centered at states $(j'_2,l'_2)$ 
complying with triangular inequality and parity conservation, take significant values. 
For the magenta, orange and green Gaussians, the edge factor $e^J_{j'_2l'_2}$ is equal to 4, 2 and 1, respectively.
A zoom of the lowest magenta Gaussian is made in order to show that exactly one fourth of the Gaussian lies 
within the yellow area allowed by the triangular inequality. \\

Fig. \ref{fig:plot4}: Rotational state distribution obtained by means of QPST and the modified CPST 
(red columns and blue diamonds, respectively). See text for more details on the calculations.  \\

Fig. \ref{fig:plot5}: Same as Fig. \ref{fig:plot1}, but with the modified QCT method. \\

\newpage

\section*{Figures}

\end{document}